\documentclass[doublecol]{epl2}
\usepackage{graphicx}
\usepackage{bm}

\title{The effect of preheating on the ablative Rayleigh-Taylor instability}
\shorttitle{Ablative Rayleigh-Taylor instability with preheating}

\author{W. H. Ye\inst{1,2} \and X. T. He\inst{1,2,3} \and W. Y. Zhang\inst{2} \and M. Y. Yu\inst{1,4}}
\shortauthor{W. H. Ye \etal}
\institute{
  \inst{1} Institute for Fusion Theory and Simulation, Department of Physics, Zhejiang University, Hangzhou 310027, P. R. China\\
  \inst{2} Institute of Applied Physics and Computational Mathematics, Beijing 100088, P. R. China\\
  \inst{3} Center for Applied Physics and Technology, Peking University, Beijing 100871, P. R. China\\
  \inst{4} Theoretical Physics I, Ruhr University, D-44789 Bochum, Germany}

\abstract{
The evolution of the ablative Rayleigh-Taylor instability (ARTI) in the
presence of preheating is investigated. When the linear ARTI growth rate has a
sharp maximum, a secondary spike-bubble structure at a harmonic of the
fundamental mode will be excited. Interaction of this secondary structure with
the latter will eventually lead to its rupture. On the other hand, for stronger
preheating the linear growth rate has no sharp maximum and in the nonlinear
evolution no secondary spike-bubble is generated. Instead, the fundamental
spike-bubble structure evolves into an elongated jet. These results are in
agreement with that from the numerical simulations and observations.
}

\begin{document}
\maketitle


During intense-laser interaction with a target in inertial confinement fusion,
a subsonic heat-conduction wave is excited at the laser-absorption spot on the
target surface. The intense heat wave propagates into the cold dense plasma and
generates a shock wave. As the latter propagates, it compresses, heats, and
accelerates the local plasma. A rarefaction wave that decelerates the
accelerating fluid is formed ahead of the heat front, where the Rayleigh-Taylor
instability (RTI) can take place \cite{1aa,Lindl,1c} 
The RTI is also common in gas combustion, the implosion phase of laser-target
interaction, as well as in astrophysical gas clouds
\cite{1aa,Lindl,1c,1a,1ab,1aaa}. 
Linear theory shows that ablation caused by the heat front can stabilize the
RTI as well as the subsequent Kelvin-Helmholtz instability
\cite{3,3aa,3b,3a,19}.

Most earlier studies of the ablative RTI (ARTI) used the fluid approach and
invoked the Spitzer-Harm (SH) conductivity for the thermal transport. On the
other hand, results from Fokker-Planck (FP) simulations showed \cite{8a,9xx}
that under ICF conditions the SH theory is inadequate because of intense plasma
heating by the laser-generated hot electrons, heat and shock waves, etc.
\cite{Lindl,1c}, and the preheated region in front of the temperature gradient
of the heat-wave front is often spatially extended. The heat transport process
involved in the region is rather complex and several models have been
introduced \cite{Lindl,1c,9xx,otani}. Simulations and analytical modeling have
shown that the preheating can significantly affect the linear growth rate and
cutoff wavelength of the ARTI, as well as its nonlinear development
\cite{8a,20,4f}. However, although the onset and initial evolution of ARTI have
been widely investigated and well understood, a more detailed physical
understanding of the later stages of the development is still lacking
\cite{1c}.

Here we consider the evolution of the ARTI accounting for the preheating effect
by analytical modeling and numerical simulation. The conservation equations
including laser absorption and electron conduction effects are solved
numerically using the LARED-S code \cite{9x}. The energy equation is given by
\cite{3,3aa,3b,9x}
\begin{equation}
\partial_tE+\nabla\cdot[(E+P)\bm{u}]=-\nabla\cdot\bm{q}(T)+S_L,
\end{equation}
where $\rho$, $\bm{u}$, $T$, $P$ and $E$ are the fluid density, velocity,
temperature, pressure, and total energy $E=\rho(\varepsilon +u^2)$,
respectively, $\varepsilon$ is the specific internal energy, $S_L$ is the
energy source due to absorption of laser energy by inverse bremsstrahlung, and
$\bm{q}=-\kappa\nabla T$ is the heat flow. To include preheating and heat flux
limitation effects, we invoke the modified SH electron heat conductivity
\cite{3,3aa,3b} $\kappa=\kappa_{\rm SH}f(T)$, where $\kappa_{\rm SH}$ is the
classical SH conductivity, and $f(T)=1+\alpha T^{-1}+\beta T^{-3/2}$ is the
preheat function \cite{1c}. Here, $T$ is in units of MK, and $\alpha$ and
$\beta$ are free parameters of order unity and can be determined empirically by
comparing with experiments, theories, and/or numerical simulations.
Thus, for constant pressure one can define a characteristic preheat length
$L=T(\partial_xT)^{-1}\sim\rho(\partial_x\rho)^{-1}=L_{\rm SH}f(T)$, where
$L_{\rm SH}=\kappa_{\rm SH}/\rho_av_ac_p$ is the classical SH length and
$\rho_a$, $v_a$, $c_p$ are the density, speed, and specific heat, respectively,
of the plasma in the ablation front region. Other heat conduction models
\cite{9b,9xx} can also be used, provided that preheating effects and heat flux
limitation are properly included.

\begin{figure}
\includegraphics[width=8cm]{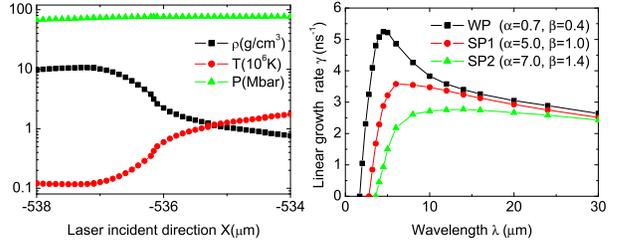}\suppressfloats
\caption{(Color online.) Left panel: profiles of the density
($0.8-10.5$ g/cm$^3$), temperature ($0.1$ -- 2 MK),
and pressure ($\sim 70$ Mb) for the WP regime at $t=9.09$ ns
near the ablation front at $x\sim -536.5 \mu$m. The laser
is incident from the right. Here electron heat conduction
heats the overdense region, and the ARTI interface is near
the density peak. Right panel:
linear (i.e., initial) growth rate of the ARTI {\it versus} the wavelength,
from numerical simulations. The black curve with squares is
for the WP case, with $\alpha =0.7$ and $\beta =0.4$. The
red curve with circles is for SP1 ($\alpha =5.0$, $\beta =1.0$),
where $\lambda_c\sim 3 \mu$m,
$L\sim 1.35 \mu$m, $v_a\sim 0.80 \mu$m/ns. The green curve with triangles
is for SP2 ($\alpha =7.0$, $\beta =1.4$.) \label{f1}}
\end{figure}

In the simulation, we first obtain a quasistationary state by irradiating a
$200 \mu$m thick CH foil target with a $0.35 \mu$m laser beam at intensity
$I=I_0\sin^4(t/5{\rm ns})/\sin^4(1.0)$ for $t<5$ ns and $I=I_0=10^{15}$W/cm$^2$
for $t\geq 5$ ns. Such a self-consistent one-dimensional quasistationary state
appears
at $t\ge 7$ ns, as shown in the left panel of fig.\ \ref{f1} for the density,
temperature, and pressure near the ablation front.
To initiate the ARTI on the ablation-front interface, we start at this instant
a small perturbation given by the fluid displacement
$x=\xi(y,0)=\delta\xi_0\cos(ky)$, where $\delta\xi_0$ is the amplitude and
$\lambda=2\pi/k$ is the wavelength. In both the dense pristine plasma and the
ablation regions, 1200 fine meshes of widths of 0.06 -- 0.07 $\mu$m are used in
the $x$ and $y$ directions. The mesh width in the $x$ direction is gradually
increased on both sides of the region. Up to $1400\times 200$ meshes are used
in the simulation.

The left panel of fig.\ \ref{f1} shows that near the ablation front the
temperature and density are $\sim 0.2$ MK and $5.6$ -- $6.5$ g/cm$^3$,
respectively, and in the ablation layer 
they are $\sim 2$ MK and $\sim 0.8$ g/cm$^3$, respectively. The right panel of
fig.\ \ref{f1} shows that the linear ARTI growth rate depends rather strongly
on the preheating length $L$. It is thus convenient to define weak preheating
(WP) by the condition $L/L_{\rm SH}=f(T)\ll f_0\sim 20$, as is the case
represented by the black squared curve (which is for $f(T)\sim 8$ at $T=0.2$
MK). Here the linear growth rate has a relatively sharp maximum at $\lambda\sim
5 \mu$m. We define strong preheating (SP) by the condition $L/L_{\rm
SH}>f_0\sim 20$. It corresponds to the regime where the linear growth rate does
not have a sharp maximum, or the red circled and green triangled curves.

\begin{figure}
\includegraphics[width=8.5cm]{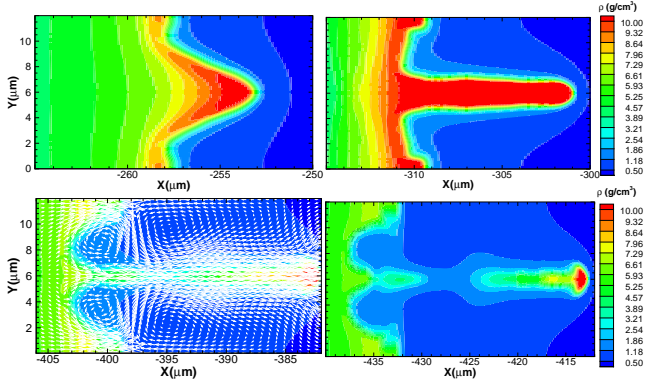}\suppressfloats
\caption{(Color online.) Amplitudes (ablation layer displacement)
of the density spikes and bubbles
for the WP regime ($\alpha=0.7$ and $\beta=0.4$).
For clarity the center of the density pattern has been moved to
$y=\lambda_1/2$. Upper left: the fundamental (or master) spike and
bubble pattern at $t=7.78$ ns. Upper right: the secondary density
spikes (around $y=0, 12 \mu$m) at $t=8.10$ ns. Lower left:
the mass flux $\rho\bm{u}$ at
$t= 8.56$ ns. Lower right: the fundamental spike
ruptures at $t= 8.73$ ns.\label{f2}}
\end{figure}

\begin{figure}
\includegraphics[width=8.5cm]{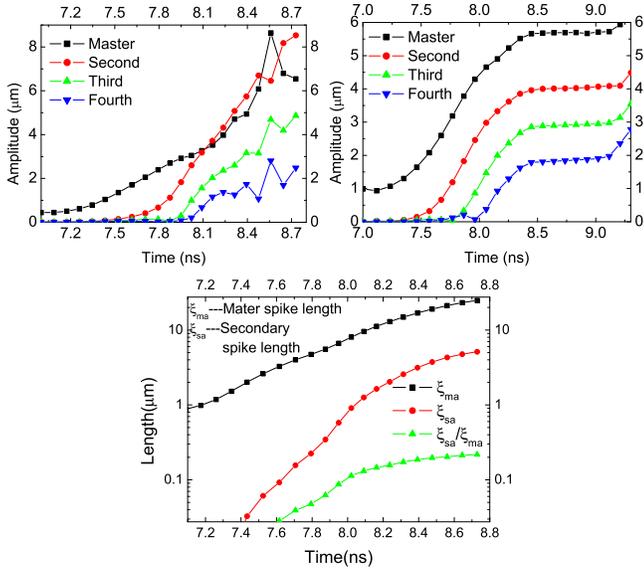}\suppressfloats
\caption{(Color online.) The amplitudes of the density
spikes {\it versus} time. Upper left: the fundamental mode and the second,
third, and forth harmonics for the WP case. Upper right:
the same for the SP1 case. Lower panel: the
lengths of the fundamental and second spikes,
and their ratio. \label{f3}}
\end{figure}

\begin{figure}
\includegraphics[width=8.5cm]{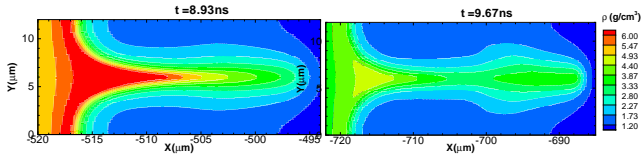}
\caption{(Color online.) Evolution of the fundamental spike and bubble
for the SP1 regime ($\alpha=5.0$, $\beta=1.0$). The initial modulation wavelength is
$12 \mu$m. Left: $t=8.93$ ns, right: $t=9.67$ ns.\label{f4}}
\end{figure}

We first consider the WP regime. For $\alpha=0.7$ and $\beta=0.4$ the
preheating length $L$ is about 0.58 $\mu$m, with $f_0\approx 8$. The initial
perturbation has $\delta\xi_0$=0.4$\mu$m and $\lambda=12\mu$m. Figure \ref{f1}
shows that the linear growth rate has a sharp maximum at $\lambda_m=0.375
\lambda_1$ and a cutoff wavelength at $\lambda_c=0.14 \lambda_1$. As expected,
the linear growth rate agrees well with that from the modified Lindl-Takabe
formula \cite{3} $\gamma=[Akg/(1+\alpha_0kL)]^{1/2}-\beta_0v_ak$, where
$A={\cal O}(1)$ is the Atwood number, $g$ is the fluid acceleration, and the
free parameters $\alpha_0$ and $\beta_0$ are in the range 2--3. That is, in the
short wavelength regime the effect of ablation is significant and the linear
growth rate decreases as the preheating length $L$ increases.

The simulation results show that besides the fundamental mode ($\lambda_1=12
\mu$m), at least six harmonics having wavelengths larger than $\lambda_c=0.14
\lambda_1$ appear. The evolution of the fundamental spike
$\xi(y-\lambda_1/2,t)$ is shown in fig.\ \ref{f2} and the evolution of the
harmonic modes $\xi_n(t)$ at $y=\lambda_1/2$ for $n=1$ -- 4 are shown in fig.\
\ref{f3}. At $t=7.78$ ns, the fundamental spike has evolved into a wide spatial
structure with a bubble of size $\sim 0.24\lambda_1$. In the plasma region that
is pre-heated by the heat conduction wave, it grows at $>10^6$ cm/sec. In this
region the temperature ($>1$ MK) and pressure ($>70$ Mb) are also high, but the
density decreases sharply, as can be seen in the left panel of fig.\ \ref{f1}.
The front of the fundamental spike becomes flattened, but no mushroom structure
is generated. This behavior can be attributed to the fact that there is
enhanced energy loss due to increased electron heat conduction, so that
velocity shear (and thus fluid rotation) is suppressed \cite{1c}.

On can see that at $t=8.1$ ns the fundamental spike is still extending forward,
but now at the slower rate $\sim 2.5\times 10^6$ cm/sec. The displacement $\xi$
of the layer is close to the modulation wavelength $\lambda_1$. That is, higher
harmonics and mode coupling effects become important, and the ARTI can be
considered to enter a new nonlinear phase. Here the boundary region of the
high-density area is gradually ablated because of the high temperature there.
In fact, the upper right panel of fig.\ \ref{f2} shows that the high-pressure
ablation layer surrounding the spike is being compressed into a narrow region
of width $\sim 2 \mu$m and density $\sim 10$ g/cm$^3$. The ablation also leads
to significant increase of the bubble pressure due to fluid expansion around
the boundaries of the region, as can be seen near the spike bottom, where the
expansion (and ablation) speed is $\sim$ 2 -- 5 $\times 10^5$ cm/sec.
Consequently, the bubbles on both sides of the spike are extended and deformed,
and the fundamental spike becomes even narrower. The bubbles are now nearly
stationary.

The upper right panel of fig.\ \ref{f2} shows that second becomes strongly
compressed. The secondary spikes can be attributed to the second harmonic
($\xi_2(t)$) having the wavelength $\lambda_1/2$. The corresponding rate of
fluid displacement can be larger than that of the fundamental spike and thus
compete for mass and energy feed with it. In fact, at $t>8.1$ ns one finds
$\xi_2(t)\geq \xi_1(t)$ in the left panel of fig.\ \ref{f3}, i.e., the
secondary spikes are overtaking the fundamental spike.

As already invoked a priori in the interpretation of the simulation results,
the evolution of the ARTI can be understood in terms of the nonlinear
interaction among the fundamental and the secondary spikes associated with the
harmonics of the fundamental excitation. For convenience, we shall adopt the
single-mode approach often used for considering the nonlinear RTI
\cite{Lindl,1c,4x,4a,4b,4c,4f,7,7a,7b}. Generation of higher harmonics in the
region of interest and their interactions shall be investigated.
An unstable initial (the fundamental) single mode perturbation with wave number
$k_1=2\pi/\lambda_1$ can excite harmonic modes with wave numbers $k_n=nk_1$,
where $n=2,3,...$ \cite{Lindl,1c,otani}. The total surface displacement of the
interface can thus be expressed as $\xi(y,t)=\sum_n\xi_n(y,t)\cos(k_ny)$. In
the weakly nonlinear stage, $\xi(y,t)$ can be represented by the first three
harmonic modes \cite{4x}. As the instability grows, more modes are included.
Accordingly, we first describe the structure of the ARTI modes by the fluid
displacement
$\xi(y-\lambda_1/2,t)=\sum_n\xi_n(y-\lambda_1/2,t)\cos[k_n(y-\lambda_1/2)]$,
where $n=1,2,3,...$ with $n=1$ being the fundamental mode (to maintain symmetry
in the analysis, in this representation the $y$ coordinate has been displaced
by $\lambda_1/2$).
Accordingly, one obtains
$\xi_{sa}(\pm\lambda_1/2,t)=$$-\xi_1(t)+\xi_2(t)-\xi_3(t)+\xi
_4-\xi_b(\pm\lambda_1/6,t)$, where $\xi_b(\pm\lambda_1/6,t)
=[\xi_1(t)-\xi_2(t)-\xi_4(t)]/2-\xi_3(t)$ is the displacement at the bottom of
the bubble, or $\xi_{sa}(\pm\lambda_1/2,t)=-3[\xi_1(t)-\xi_2(t)-\xi_4(t)]/2$.
Since $\xi_1\sim \xi_2$, we have $\xi_{sa}(\lambda_1/2,t)\sim 3\xi_4(t)/2>0$.

In fig.\ \ref{f3}, the profiles of $\xi_n(t)$, $n=1$ -- 4, obtained by
numerically solving the governing mode equations are shown. Using the value of
$\xi_4$ at $t=8.1$ ns from the upper-left panel, we find $\xi_{sa}=1.2 \mu$m
(for the same instant) from the lower panel. This result agrees well with the
corresponding value $1.3 \mu$m (for the secondary spikes in fig.\ \ref{f2},
upper right panel) from the simulations. That is, by including four harmonics
in the theory the obtained maximum displacement $\xi_{sa}$ of the secondary
spikes agrees with that of the simulations.

For $\xi_{sa}<0$ and $\xi_1>\xi_2> ...$, the distinct maximum in the linear
ARTI growth rate becomes smoothed out, and there is no secondary spike
excitation. Instead, at $t = 8.1$ ns new cavitation (see fig.\ \ref{f2},
upper-right panel) near the secondary spikes appear. The cavitation and the
fundamental bubble coalesce into an irregular, transversely widened, large
bubble between the fundamental and the secondary spikes. The fundamental spike
is further compressed and becomes still narrower. At $t = 8.56$ ns, new eddies
of the mass flux are generated between the fundamental and secondary spikes, as
indicated by arrows in the lower-left panel of fig.\ \ref{f2}. The mass flux
pushes against the waist of the fundamental spike, such that the density of
upper region of the fundamental spike increases to $\rho>7.5$ -- 10 g/cm$^3$
and that of the lower part is less. The upper part is accelerated upward and
becomes a jet. The secondary spike grows faster than the fundamental one. The
ratio $R=\xi_{sa}(\pm \lambda_1/6,t)/\xi_{ma}(0,t)$, where
$\xi_{ma}(0,t)=\sum_n\xi_n(t)$ ($n=$1 -- 4), is a measure of the displacement
of the fundamental spike-bubble pattern at the center. It
increases gradually, as shown by the red and green curves in the lower panel of
fig.\ \ref{f3}. That is, the secondary spikes acquire more fluid mass than the
fundamental spike. The plasma at the foot of the fundamental spike also feeds
the secondary spikes, making them grow faster, as shown in the lower left panel
of fig.\ \ref{f2}, where the arrows indicate the direction and magnitude of the
mass flux.
Eventually, at $t = 8.73$ ns the fundamental spike ruptures into two parts, as
observed in the simulation result (fig.\ \ref{f2}, lower right panel).
Simulations also show that for the parameters considered the threshold above
which the fundamental spike will eventually rupture is $R\sim 0.20$.

If the initial modulation wavelength is larger than $\lambda_1$=12 $\mu$m, we
found by including in the analysis harmonics higher than the fourth that at
least 14 higher harmonics are excited and the first 12 modes have initial
growth rates larger than that of the fundamental mode. That is, many different
modes can be excited simultaneously and the long-wavelength fundamental mode in
our model is not preferred (even though it is started with a finite amplitude).
The overall grow rate is less than the preceding cases and the evolution
behavior of the SP and WP regimes can no longer be distinguished, as can also
be seen from the linear growth-rate {\it versus} wavelength curves (see fig.\
\ref{f1}, right panel). In fact, the ARTI growth rates for longer wavelength
perturbations varies slowly with the wavelength. In the simulation result we
see that two other spikes (corresponding to the additional shorter wavelength
perturbations) appear in the original cavity region. The fundamental spike
eventually ruptures due to lack of mass feed resulting from competition for the
latter among the three spikes.

As can also be seen in the right panel of fig.\ \ref{f1}, the linear growth
rate in the SP regime is smaller than that of the WP regime, and it has no
sharp maximum. This is because the SP regime corresponds to longer preheating
length $L$ and lower Atwood number. For the SP case of $\alpha=5.0$ and
$\beta=1.0$ (denoted by SP1), the initial perturbation and wavelength remain
the same as that without preheating. The cutoff wavelength is about $2.8 \mu$m
and the linear growth rate has no sharp maximum. For the $\alpha=7.0$ and
$\beta=1.4$ case (denoted by SP2), the cutoff wavelength is about $3.6 \mu$m.
The growth rate versus wavelength curve is relatively smooth and the local
maximum is not sharp. Examining the results of the SP1 case in more detail, we
found that except in the early linear growth phase, no harmonic mode grows
faster than the fundamental mode, as shown in the right panel of fig.\
\ref{f3}. This can be attributed to the fact that the input energy is spread
over several modes, instead of concentrating only in the mode with the highest
(sharply peaked) growth rate, as in the WP case. In fact, in the SP regime the
growth of the fundamental and the secondary harmonics in the nonlinear stage is
similar. As a result, the fundamental spike does not rupture. Instead, it
evolves into a long jet that spurts towards the higher-temperature underdense
plasma at an average speed of $\sim 1.8\times 10^6$ cm/sec, as shown in fig.\
\ref{f4}. Such long jets have been frequently observed in experiments
\cite{1c,10a} as well as astrophysical clouds \cite{1,1a}. However, it should
be cautioned that when several harmonics are excited simultaneously, the
single-mode assumption can become invalid and other modes that are not
harmonics of the fundamental mode can also be involved. In this case turbulent
mixing is expected to dominate.


In summary, we have shown that the evolution of ARTI in the presence of
preheating can be interpreted in terms of the generation and interaction of
harmonics of the fundamental perturbation. In agreement with that from existing
numerical simulations \cite{4c,20,4f}, it is found that the initial growth of
the fundamental spike-bubble system depends on the ablation length. However,
the evolution of the ARTI also depends significantly on the preheating
parameters. In fact, there can be two regimes of ARTI evolution. The WP regime
is characterized by a distinct peak in the wavelength dependence of the linear
growth rate and rapid excitation of the corresponding, namely the second,
harmonic. In the nonlinear stage of the evolution of the instability the
fundamental and the second harmonic modes compete with each other. Eventually
the second harmonic dominates and the fundamental spike ruptures. In the SP
regime, on the other hand, the linear growth is weaker and not sharply peaked,
so that many harmonics are excited. In the nonlinear evolution the fundamental
spike remains dominant and evolves into a long jet that spurts towards the
higher-temperature underdense blow-off region \cite{Lindl,1c}. Both scenarios
of the ARTI evolution have often been observed in the laser-plasma compression
experiments as well as in astrophysical plasmas \cite{1c,1a,1aa,1ab,7b,10a}.

\acknowledgments
We thank G. M. Lu and L. F. Wang for useful discussions. This
work was supported by the National High-Tech ICF Committee, the National
Natural Science Foundation of China (10775020, 10835003, and 10935003), and the
National Basic Research Program of China (2007CB815100, 2007CB814802,
2008CB717806) and the Ministry of Science and Technology ITER Program
(2009GB105005).


\suppressfloats

\end{document}